
\documentclass[twocolumn]{aastex63}
\usepackage{epsfig}                     
\usepackage{graphicx}                    
\usepackage{courier}     
\usepackage{natbib}                 
\usepackage{amssymb}                    
\usepackage{color}                       
\usepackage{hyperref}
\usepackage[space]{grffile}

\graphicspath{{./}{figures/}}

\begin{document}

\title{Prompt emission of relativistic protons up to GeV energies from M6.4-class solar flare on July 17, 2023}

\author{C.E. Navia}
\affiliation{Instituto de F\'{i}sica, Universidade Federal Fluminense, 24210-346, Niter\'{o}i, RJ, Brazil }
\correspondingauthor{C.E. Navia}
\email{carlos\_navia@id.uff.br}

\author{M.N. de Oliveira}
\affiliation{Instituto de F\'{i}sica, Universidade Federal Fluminense, 24210-346, Niter\'{o}i, RJ, Brazil }


\author{A.A. Nepomuceno}
\affiliation{Departamento de Ci\^encias da Natureza, Universidade Federal Fluminense, 28890-000, Rio das Ostras, RJ, Brazil}

\begin{abstract}
We show evidence of particle acceleration at GEV energies associated directly with  protons from the prompt emission of a long-duration M6-class solar flare on July 17, 2023, rather than from protons acceleration by shocks from its associated Coronal Mass Ejection (CME), which erupted with a speed of 1342 km/s. Solar Energetic Particles (SEP) accelerated by the blast have reached Earth, up to an almost S3 (strong) category of a radiation storm on the NOAA scale.  
Also, we show a temporal correlation between the fast rising of GOES-16 proton and muon excess at ground level in the count rate of the New-Tupi muon detector at the central SAA region.
A Monte Carlo spectral analysis based on muon excess at New-Tupi is consistent with the acceleration of electrons and protons (ions) up to relativistic energies (GeV energy range)  in the impulsive phase of the flare. In addition, we present another two marginal particle excesses (with low confidence) at ground-level detectors in correlation with the solar flare prompt emission.
\end{abstract}


\section{Introduction} 
\label{sec1}

Since 1950 the observation of solar energetic particles from the solar 
flares and coronal mass ejections (CMEs) have been done with ground-level experiments, such as the neutron monitors (NMs) \citep{meye56,simp00,mora00}
as well as the solar neutron telescope network \citep{hu22,vald09}, all around the world.
These observations have yielded a lot of new information. For instance, the existence of a prompt and gradual emission of solar energetic particles (SEP) in flares and CMEs, respectively, the correlations of the cosmic ray intensity with CMEs and other solar disturbances crossing the Earth, etc.
\citep{chup87,mora00}

 Also, the solar modulation of galactic cosmic rays is inversely correlated with solar activity, inferred through the number of sunspots, which can be the key to understanding more about space weather \citep{cade15}.

Nowadays, particles accelerated to near the Sun can be detected by space-borne instruments such as the High-Energy Proton
and Alpha Detector (HEPAD) on the Geostationary Operations Environmental Satellite (GOES) and the  Advanced Composition Explorer (ACE) spacecraft at Lagrange L1 point, through the Electron Proton Alpha Monitor (EPAM) and the Solar Isotope Spectrometer (SIS), among others.

Not all of the solar energetic particles can be measured at ground level. Even those SEPs from solar events with a good geoeffectiveness can be dissipated by the IMF, or deflected or captured by the Earth's magnetic field or until absorbed by atmosphere.
 
 On the other hand, ground-level enhancements (GLEs), typically in the MeV-GeV energy range, are sudden increases in cosmic ray intensities registered in most cases by NMs. GLEs are quite rare events, and fewer than 100 GLEs have been observed by NMs in the last 70 years. In most cases, the NMs that observed GLEs are located at regions with small geomagnetic rigidity cutoff, that is, at high latitudes \citep{shea12}.
 
\begin{figure*}[th]
\vspace*{-0.0cm}
\hspace*{-0.0cm}
\centering
\includegraphics[clip,width=1.0
\textwidth,height=0.5\textheight,angle=0.] {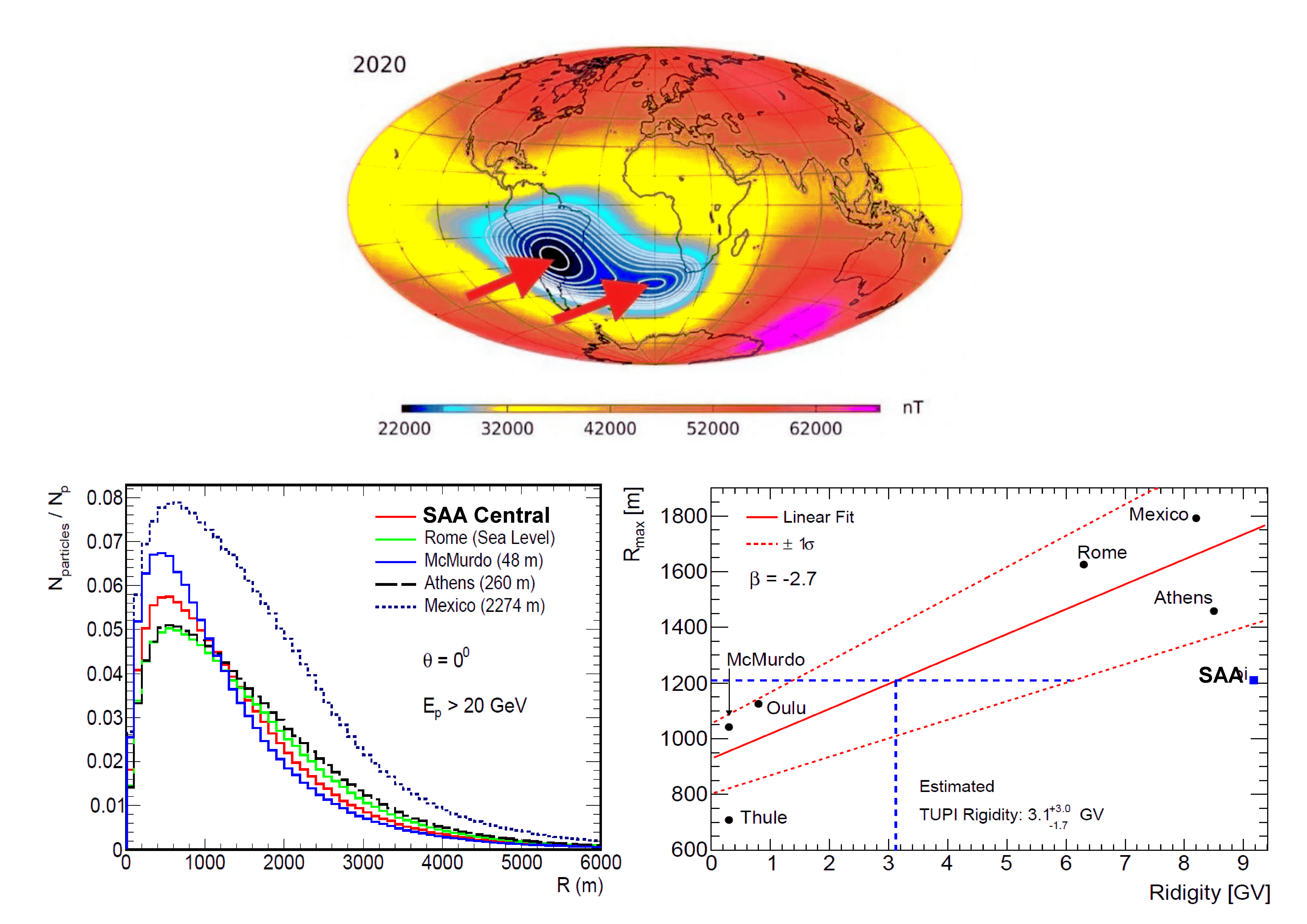}
\vspace*{-0.0cm}
\caption{Top panel: strength of Earth's magnetic field in 2000, according to measurements
by the European Spacing Agency's SWARM satellites.
Bottom left panel: Corsika-Fluka simulation results of the lateral particle distribution in proton air-showers of cosmic rays, 
as detected from several ground level detectors. Bottom right panel: correlation between $R_{max}$ (from Corsika-Fluka) versus the 
geomagnetic Stormer rigidity cutoff of six different places (black circles) including the SAA-CR region (blue square). The red solid line is a linear fit, and the two dotted red lines delimit the region 
with a confidence of 
$\pm 1 \sigma$.  }
\label{saa_tupi}
\end{figure*}

 The GLEs follow the solar radiation storms, solar energetic particles (mostly protons) observed by GOES. They occur when a large-scale magnetic eruption, a coronal mass ejection and associated solar flare, accelerates charged particles in the solar atmosphere to high energies.
 
However, in the present case, despite a radiation storm reaching above the S2-class on the NOAA scale on July 18, 2023, it did not generate a GLE, only a prompt emission of relativistic protons (ions) above GeV energies, during the phase eruptive,  and observed by ground-level detectors strategically located, within the SAA central region (New-Tupi muon detector) and by the (Yangbagin muon telescope) at the Yangbajing Cosmic Ray Observatory (Tibet 4440 m a.s.l) \citep{zhan10}.

Also, we looked for any signal in the counting rate at the Neutron Monitor's (NM) network around the world from  Neutron Monitor Data Base (NMDB)
\url{https://www.nmdb.eu/nest/}, with negative results. However, we found a low confidence signal only at Kerguelen NM, at geographical coordinates (49.3S, 70.3E), altitude of 33 m a.s.l, and an effective vertical cutoff rigidity of 1.14 GV.
We present details of these observations.

\section{New-Tupi telescope within the South Atlantic Anomaly}

The New-Tupi muon detector is completely
unmoderated (without no surrounding lead or other material). The muon detection energy threshold is about 200 MeV (see Appendix A). That contrasts with other muon detectors that have, in most cases, a surrounding lead material with a  thickness of up to 5 cm.

The shielding effect of the Earth's magnetic field on cosmic ray particles is quantified by the magnetic rigidity cutoff from a specific location \citep{smar09}. The smaller the rigidity cutoff, the lower the energy cosmic ray particles penetrate the magnetosphere.
On the other hand, a restricted area between latitudes 20 and 40 of the southern hemisphere, over South America
and the Atlantic Ocean poses a geomagnetic field
with an anomalously lower intensity (around 22,000
nT). The region is known as the South Atlantic Anomaly
(SAA) \citep{pavo16}. According to Swarm's satellite observations
\citep{finl20}, the SAA appears splitting into two, a smaller area
over the Atlantic Ocean in southwest Africa and a larger
area over eastern South America. Fig.~\ref{saa_tupi} (top panel) summarizes
the situation. We would like to point out that
the location of the New-Tupi telescope coincides with the central
part of the SAA indicated by the arrow on the left
of Fig.~\ref{saa_tupi} (top panel).

The main effect of the SAA is on the satellites since
the '70s. We know the frequent failures when they pass
through the SAA region. A large amount of charged
particles precipitation in this region damages and perturbates
the satellites' electronics. Also, according to
the results from the PAMELA detector at satellite's altitudes
\citep{caso09}, the effect of geomagnetic cutoff on low-energy
particles is present in high latitudes close to the poles
and also in the SAA region, composed mostly of low energy
cosmic protons (E $<$ 200 MeV ). In other words, the
Pamella satellite has shown that the SAA introduces a
sub-cutoff in the magnetic rigidity, below the Stormer's
magnetic rigidity cutoff.
We show that the SAA also affects secondary cosmic
rays detected at ground level. As the horizontal magnetic
component on Earth's surface is smaller on the
SAA, the magnetic lateral dispersion of the secondary
particles forming an air shower is smaller too. The effect
increases the number of particles reaching a detector. In
other words, this behavior mimics a magnetic rigidity
sub-cutoff below the Stormer's rigidity cutoff.
We show that effect through a Monte Carlo simulation
based on CORSIKA-Fluka code \citep{heck12,batt08}, where $1.0 \times 10^6$ proton
air-showers are simulated, taking into account the magnetic
coordinates (latitude, longitude) and height of several
places where detectors are installed (mostly neutron
monitors).

Fig.~\ref{saa_tupi} bottom left panel shows the lateral particle distribution
in air-showers of cosmic rays, as detected from
several ground-level detectors. In all cases, there is a fast
rise of particles with the shower lateral development until
reach um maximum value that happens for different
values of R, called hereafter as $R_{max}$
We can see that the number of shower particles at
$R_{max}$ in the SAA central region (SAA-CR) rigidity 9.6
GV is higher than at Rome and Athens, both with the
rigidity of 6.3 GV, and 8.5 GV, respectively, i.e., minors
than the SAA-CR.
Already Fig. 1 bottom right panel, shows a correlation
between $R_{max}$ versus the geomagnetic Stormer rigidity
cutoff of six different places (black circles), including the
SAA-CR (blue square). The solid red line is a linear fit, and the two dotted red lines delimit the region with significance of $\pm 1 \sigma$.
Only two places are out from the $\pm 1.0 \sigma$ significance region, the Thule (Groenlandia) in the lowest rigidity region
and SAA-CR in the highest rigidity region. The
high Stormer's rigidity of SAA-CR does not correspond
to the high value of $R_{max}$ expected by the correlation.
From an interpolation, it is possible to see that the small
value of $R_{max}$ at SAA-CR correspond to the rigidity of
only $3.1^{+3.0}_{-1.7}$ GV, within a confidence of $\pm 1.0 \sigma$.
This behavior of having a location close to the Equator,
with a nominal lower magnetic rigidity cutoff, favors the
observation of phenomena such as the SEPs.

\section{Analysis}

\begin{figure}[]
\vspace*{-0.7cm}
\hspace*{-1.0cm}
\centering
\includegraphics[clip,width=0.56
\textwidth,height=0.80\textheight,angle=0.] {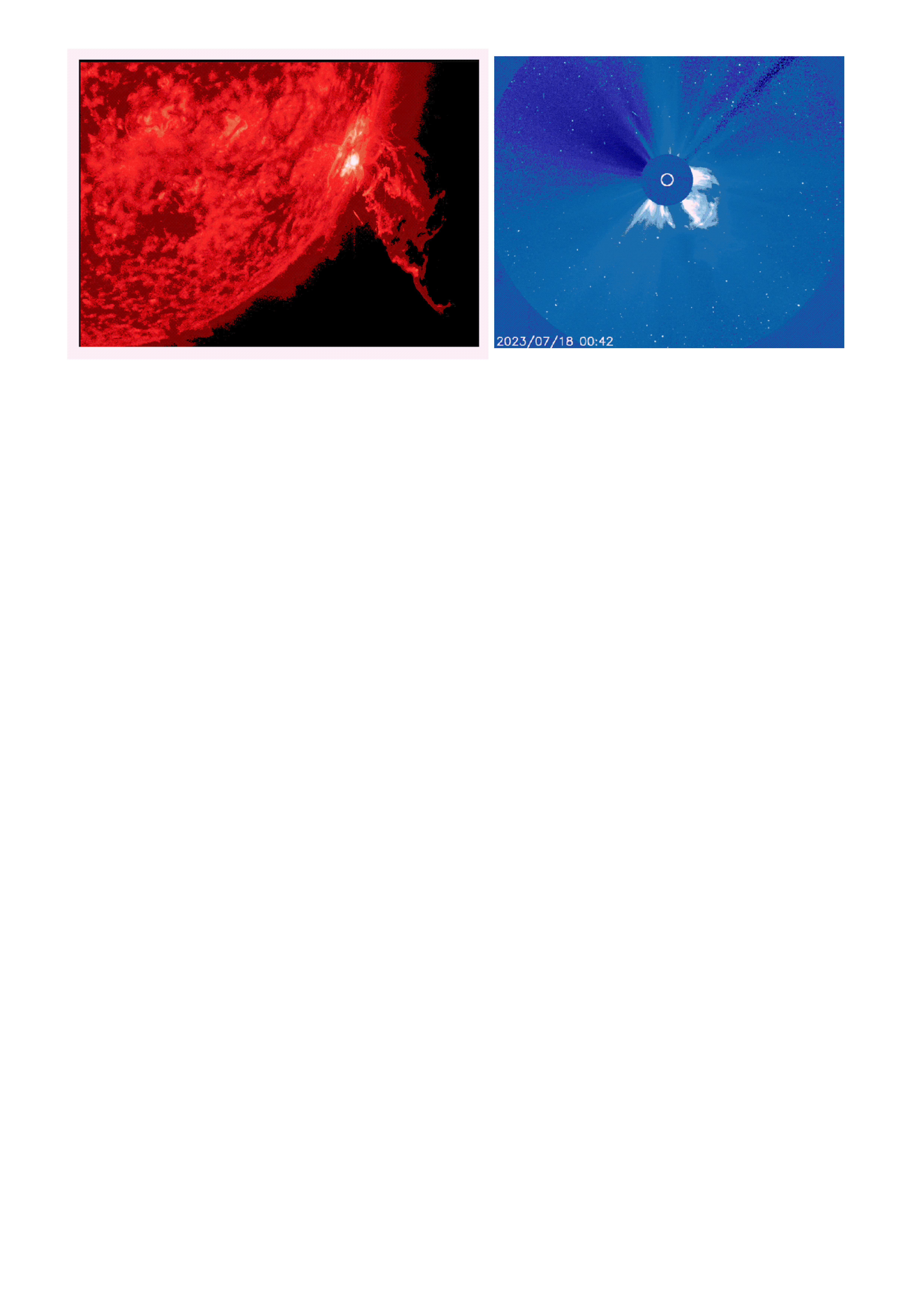}
\vspace*{-13.5cm}
\caption{Left panel:  NASA's Solar Dynamics Observatory image of a blast near the sun's southwestern limb, erupted from a big sunspot AR3363, with onset at the last hours of July 17, 2023.
Right panel: LASCO-C2 coronograph image on July 18, 2023, at 
00:42 UT showing the CME eruption associated to M6-class flare with a speed of 1342 km/s. 
}
\label{flare_cme}
\end{figure}

On July 17, 2023, at $\sim$18h UT, the active region AR 13363 had an explosion, reaching an M6-class solar flare followed by a resplendent coronal mass ejection. Fig.~\ref{flare_cme} left panel shows the image from the Solar Dynamo observatory of the blaze of fire responsible for the X-ray flux reaching M6-class flare. Already the right panel shows the LASCO-C2 coronograph image of its associated CME on July 18, 2023, at 00:42 UT.

NOAA prediction models confirmed that a CME originated in the powerful M6-class flare from sunspot AR3363 would pass through the magnetosphere on July 20, triggering at least a G1-class (minor) geomagnetic storm. However, no magnetic storms were observed. 

\begin{figure}[]
\vspace*{-1.0cm}
\hspace*{-0.5cm}
\centering
\includegraphics[clip,width=0.5
\textwidth,height=0.4\textheight,angle=0.] {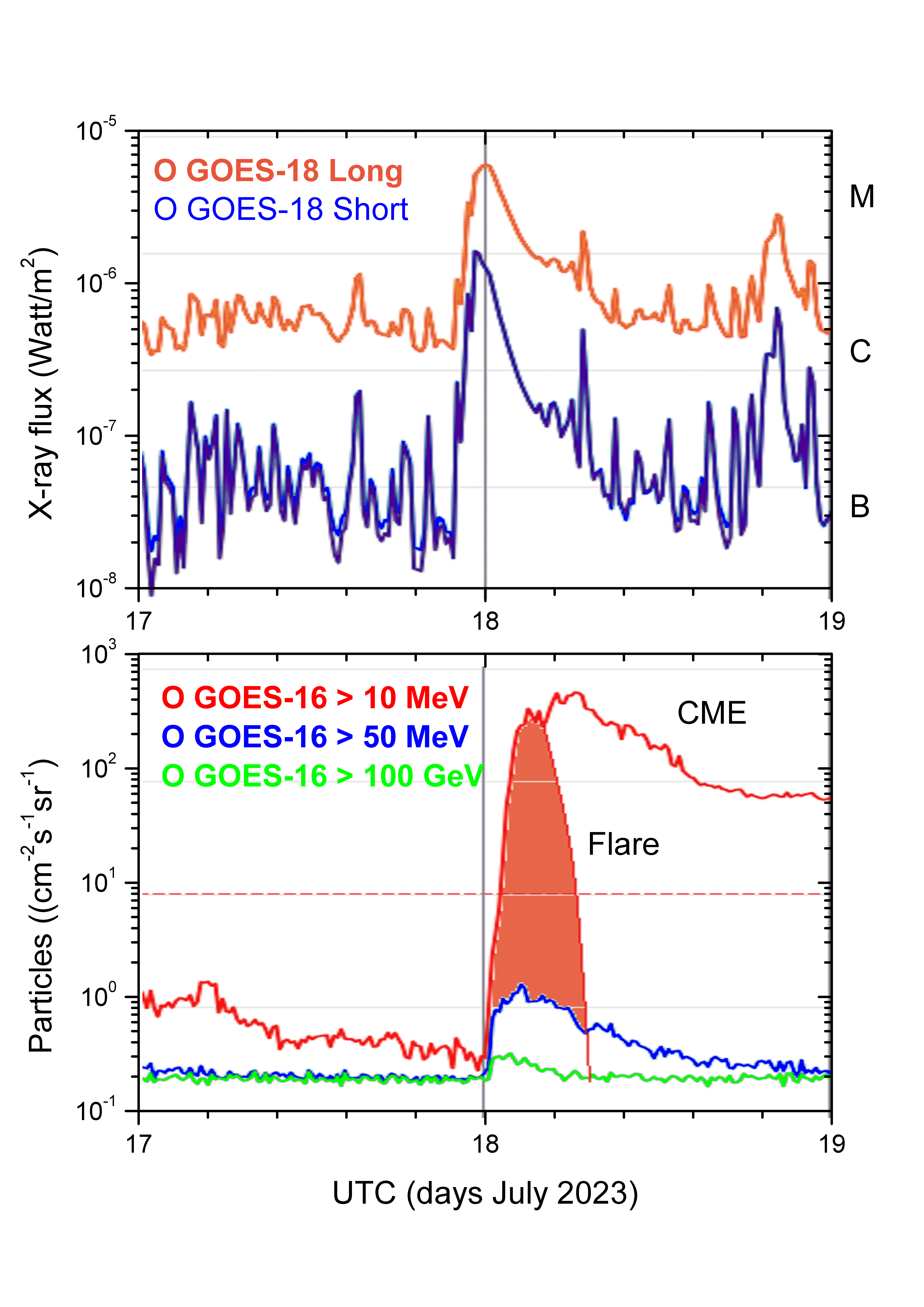}
\vspace*{-0.5cm}
\caption{Top panel: GOES-18 X-ray flux in two wavelengths.
Bottom panel: GOES-16 proton flux in three energy bands. Both on July 17-18, 2023. The orange area at the bottom (is a visual guide only) highlights the proton prompt emission during the flare impulsive phase.
}
\label{goes_goes}
\end{figure}

Fig.~\ref{goes_goes} shows the GOES-18 X-ray flux (upper panel) and the GOES-16 proton flux (bottom panel). The X-ray flux peaks at 18:00 UT, while the proton flux has two peaks. The first (in orange) is due to the acceleration of protons during the impulsive fast-rising phase of the flare peaking at 18:09 UT. The delay between the X-ray and proton flux peaks is because the proton velocity is slightly less than c, and the proton path is longer. The second peak are the protons accelerated by CME shocks, peaking at 18:14 UT, and it's the so-called gradual phase and is characterized by its long duration, up to several days.

\begin{figure}[]
\vspace*{-1.0cm}
\hspace*{-0.5cm}
\centering
\includegraphics[clip,width=0.5
\textwidth,height=0.4\textheight,angle=0.] {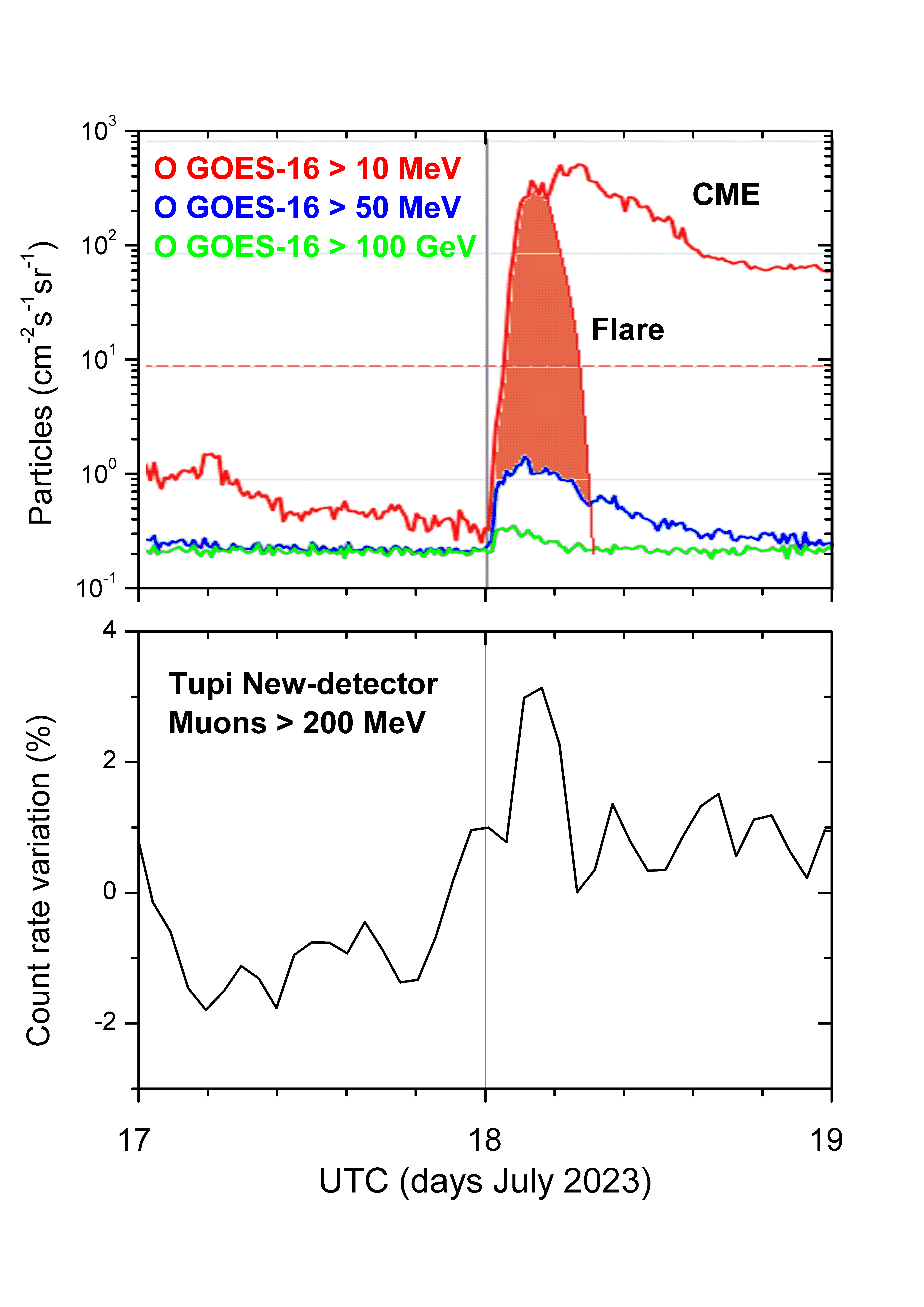}
\vspace*{-0.5cm}
\caption{Top panel: GOES-16 proton flux in three energy bands. Bottom panel: New-Tupi counting rate expressed in variation (\%). Both on July 17-18, 2023. The orange area at the top (is a visual guide only) highlights the proton prompt emission during the flare impulsive phase.
}
\label{goes_tupi}
\end{figure} 


Fig.~\ref{goes_tupi} shows the temporal coincidence between the GOES proton flux in the impulsive phase and New-Tupi muon excess. Particles (mostly proton) are accelerated in this phase exclusively by the flare, during the fast-rising until to reach the first peak (orange sector) in Fig.~\ref{goes_tupi}.

However, in the so-called gradual phase, protons accelerated by CME's shocks, the proton flux does not reach the GeV energy range because there are no excess muons at ground level.

\begin{figure}[]
\vspace*{-0.5cm}
\hspace*{-0.0cm}
\centering
\includegraphics[clip,width=0.5
\textwidth,height=0.4\textheight,angle=0.] {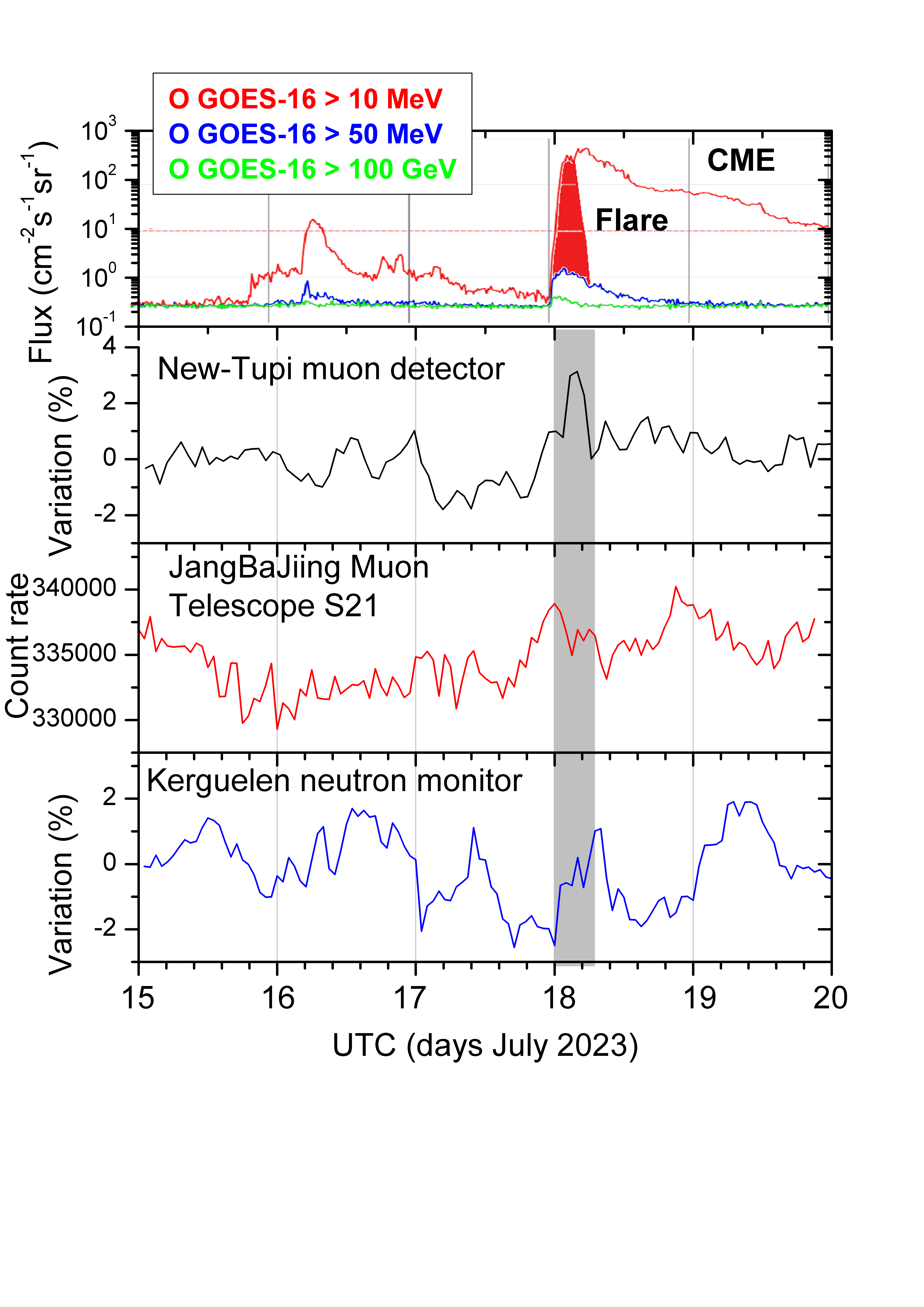}
\vspace*{-1.5cm}
\caption{From top to bottom: GOES-16 proton flux in three energy bands, 2nd to 4th panels, counting rate at New-Tupi muon detector, Yan ba Jing muon telescope, and Kerguelen Neutron Monitor, respectively.
To five consecutive days, from July 15-19, 2023.
}
\label{quatro}
\end{figure}


\section{Spectral analysis}

We perform a Monte Carlo simulation of air showers initiated by SEP (protons) using the CORSIKA code \citep{heck12}, together with the FLUKA interaction model \citep{batt08}, that works well at GeV and sub-GeV energies, including secondary particle decay. The surviving particles are tracked through the atmosphere until they reach ground level (sea level). Most particles are muons with a small contribution of electrons and nucleons.
\begin{figure}[]
\vspace*{-0.0cm}
\hspace*{-0.5cm}
\centering
\includegraphics[clip,width=0.5
\textwidth,height=0.6\textheight,angle=0.] {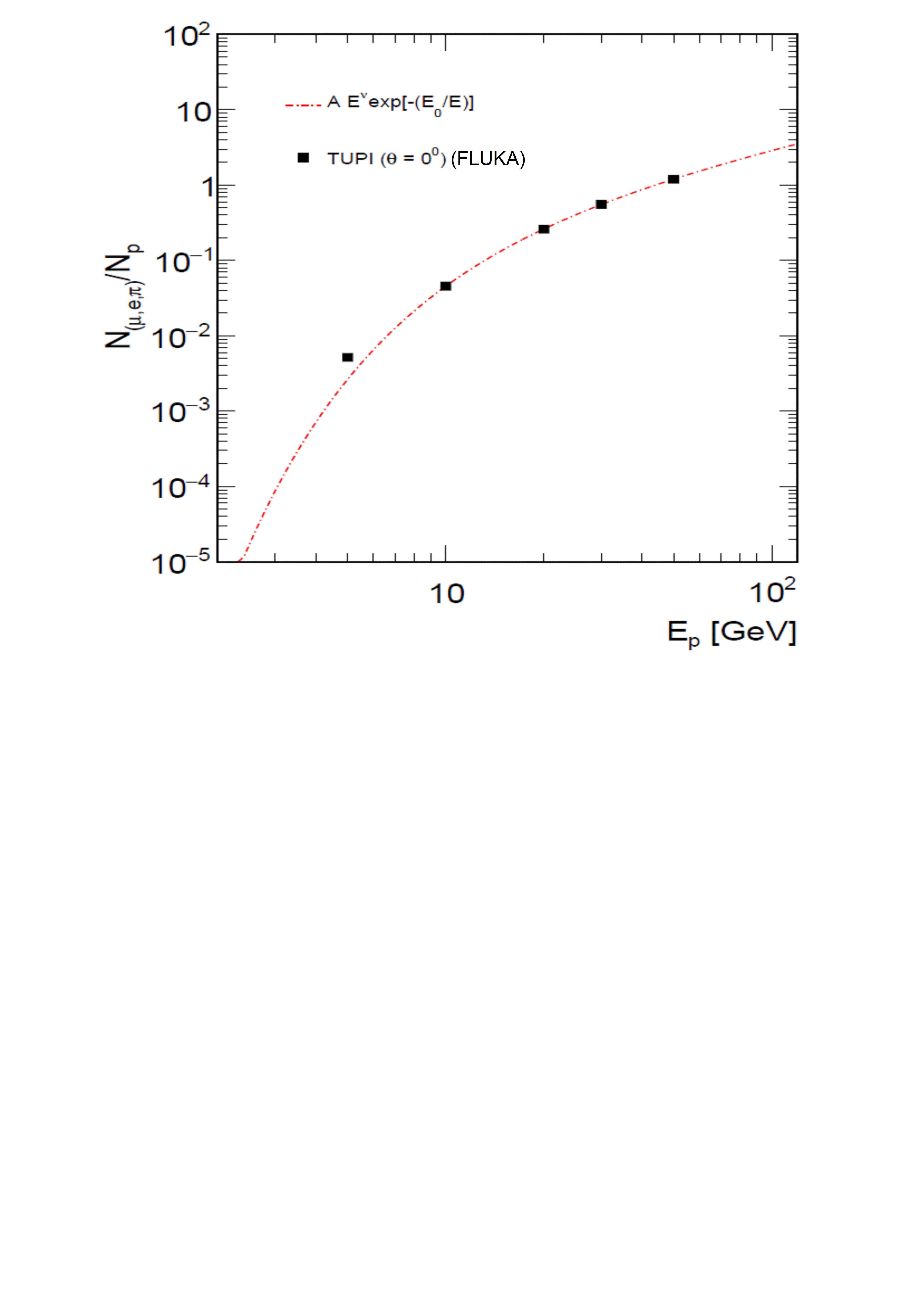}
\vspace*{-7.0cm}
\caption{Yield function, as the number of muons at the sea level per
proton (vertical incidence), as a function of incident proton energy, from CORSIKA-FLUKA simulations, taking into account the SAA’s central region magnetic conditions.
The red dashed curve shows a fit function. 
}
\label{yield2}
\end{figure} 

The aim is to obtain the yield function, $S_{\mu}(E_P)$, that is, the number of muons at sea level per primary proton, for an estimate of the upper limit of the integral proton flux in the GeV energy range, associated with the impulsive phase of M-6-class flare with onset on July 17, 2023, at $\sim $ 18 UT. 

\begin{figure}[]
\vspace*{-1.0cm}
\hspace*{-1.5cm}
\centering
\includegraphics[clip,width=0.55
\textwidth,height=0.4\textheight,angle=0.] {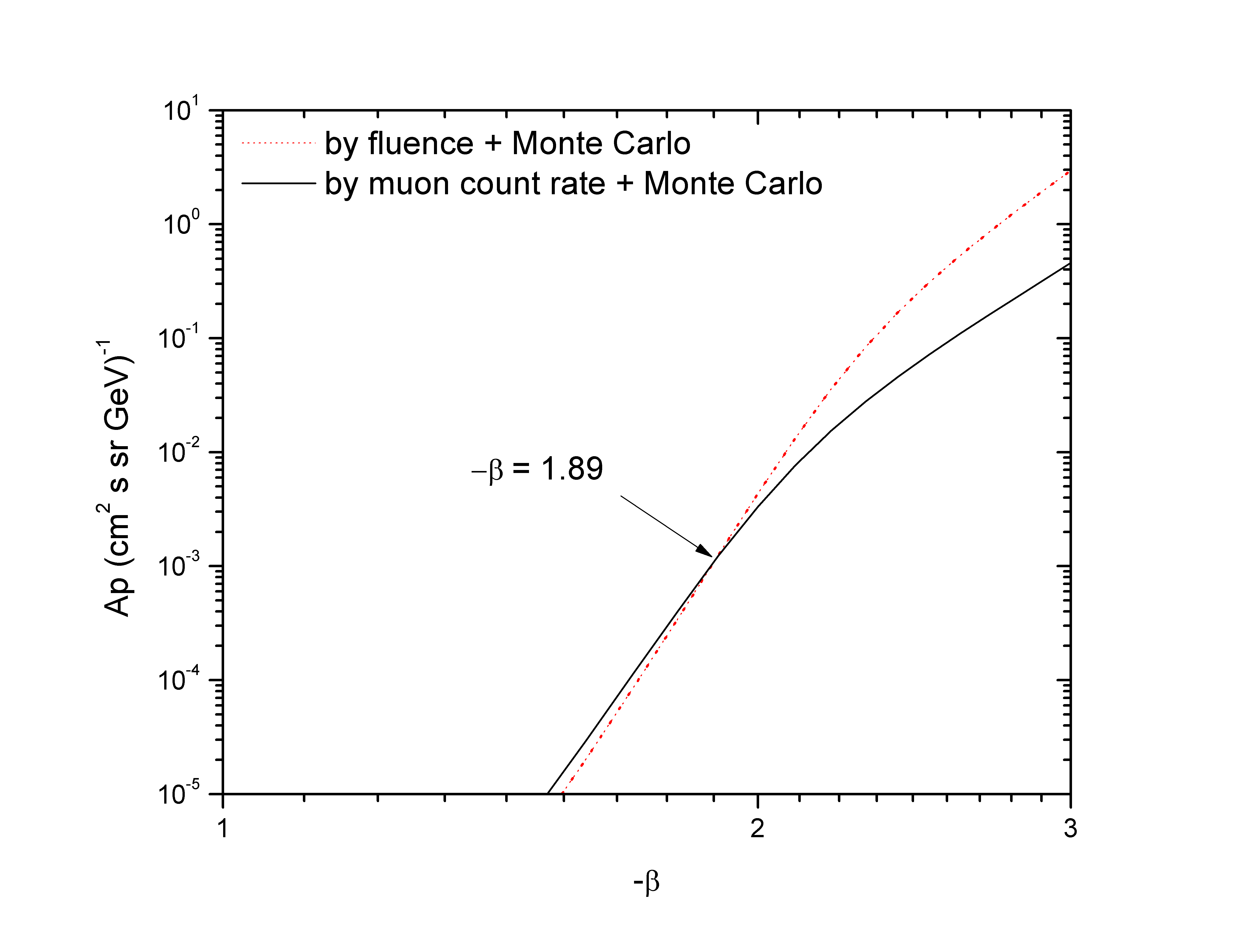}
\vspace*{-1.0cm}
\caption{Correlation between the coefficient Ap and the spectral index $\beta$. All possible values of Ap and $\beta$ compatible with the New-Tupi muon counting rate excess
(black solid curve) and the high-energy GOES-16 proton fluence F (red dot curve). These quantities are defined by equations \ref{counting} and \ref{fluence}, respectively. 
}
\label{cruze}
\end{figure} 

Fig.~\ref{yield2} (black squares) shows the Monte Carlo output under the New-Tupi geomagnetic conditions and vertical proton incidence, and fitting as 

\begin{equation}
 S_{\mu}(E_P)=A_{\mu} E_P^{\nu}exp\left(-E_{0}/E_P \right),
 \label{yield}
\end{equation}

where $A_\mu = (6.8 \pm 1.4) \times 10^{-3}$, $\nu=1.18 \pm 0.24$, and $E_0=10.2 \pm 2.1$ GeV.  Fig.~\ref{yield2} shows the fits (red dot line).

\begin{figure}[]
\vspace*{-1.0cm}
\hspace*{-1.5cm}
\centering
\includegraphics[clip,width=0.6
\textwidth,height=0.7\textheight,angle=0.] {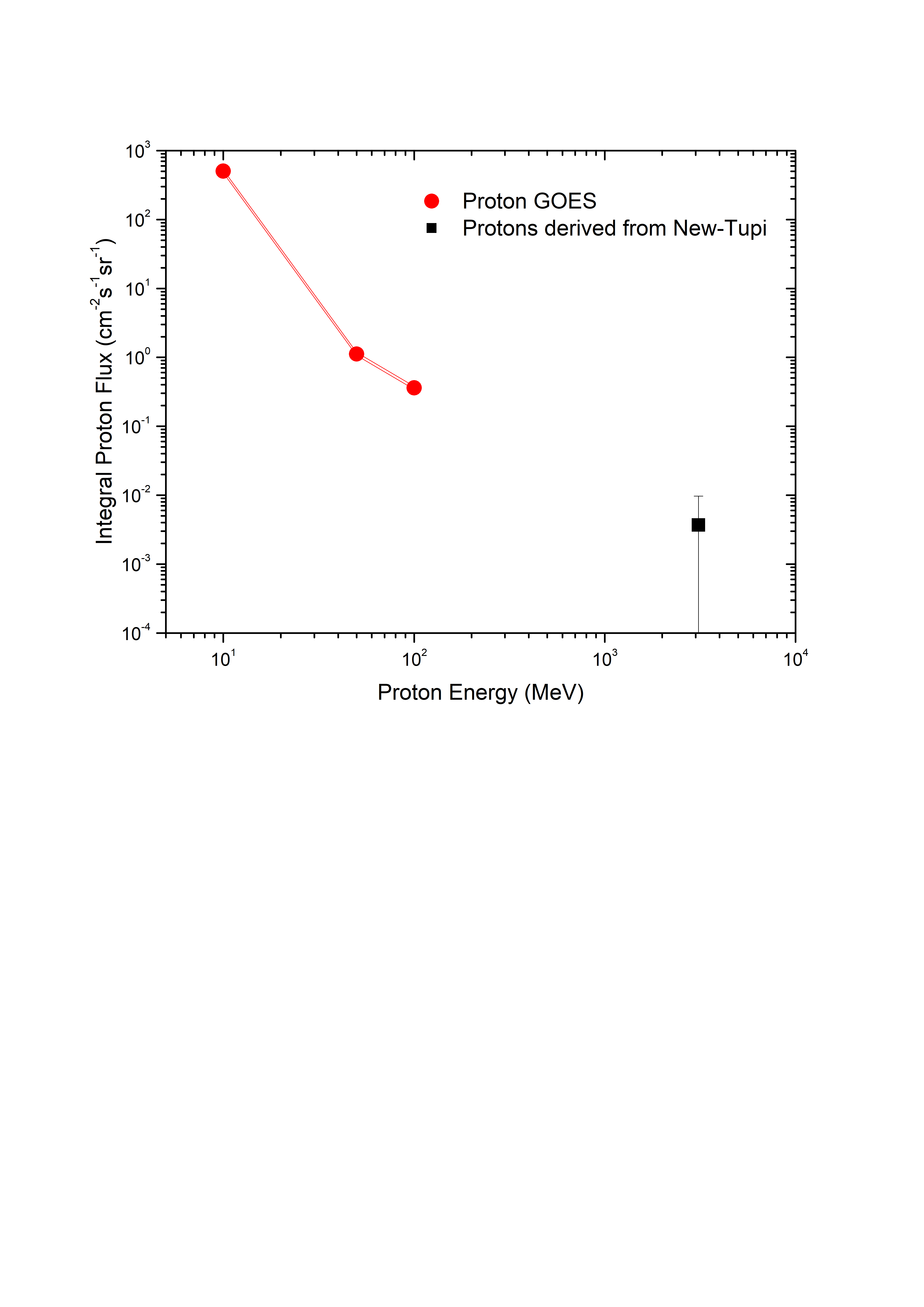}
\vspace*{-7.0cm}
\caption{Integral proton flux: the red circles represent the GOES-16 data. It corresponds to the GOES-16 proton flux prompt emission  (orange area in Fig. \ref{goes_goes}.
 The black square represents the proton flux obtained from
the muon excess on the New-Tupi detector (and Monte Carlo calculations), observed in coincidence with
the radiation storm. 
}
\label{tupi_flux}
\end{figure} 

In addition, we assume here that the energy spectrum of solar protons in the GeV energy range, which is in the high-energy tail of the SEP spectrum, can be fitted by a single power-law function.

\begin{equation}
J_P(E_P)=A_PE_P^{\beta}.
\label{power}
\end{equation} 
There are two unknown quantities in the above power-law function: the coefficient $A_P$ and the spectral index $\beta$.

A convolution between the yield function $S_{\mu}(E_P)$  and the proton spectrum $J_P(E_P)$ gives the response function \citep{augu16b}, which is the number of muons in the excess signal at New-Tupi detector generated by the SEP during the period T. We express this convolution as

\begin{equation}
J_{\mu}= \int_{E_{min}}^{\infty}S_{\mu}(E_P)F(\theta)A_P E_P^{\beta}dE_P,
\label{counting}
\end{equation}

where $F(\theta) \sim \exp(\theta/C)$ is the pitch angle distribution \citep{shea82,miro05}. In the central region of SAA (New-Tupi), the transverse geomagnetic component is only $B_{\perp} =18.3$ mT, that is, almost 80\% smaller than the transverse component, at the same latitude, but outside from the SAA region, favoring the focusing factor of the geomagnetic parallel geomagnetic on the incident solar protons (small pitch angles). 
For the present event, we found $\exp(\theta /C) \sim 1$.

The muon excess associated with protons emitted during the impulsive phase (see Fig.~\ref{goes_tupi}), and considering an effective angular aperture of  60 degrees around the zenith of the New-Tupi detector, the counting rate excess is
\begin{equation}
J_{\mu}= (3.20 \pm 2.51) \times 10^{-3} muons/(cm^2\;s\;sr),
\end{equation}

Furthermore, we also obtain the integrated time primary 
fluence as
\begin{equation}
F=T 
\left[\int_{E_{min}}^{\infty}dE_P A_P E_P^{\beta}E_P\right].
\label{fluence}
\end{equation}

For the present case, the GOES-proton fluence in the high-energy region (Ep$\geq$50) MeV is
\begin{equation}
F \sim (46.0 \pm 34.9) \;MeV/(cm^2 sr),
\end{equation}

The terms on the left side of Eq.~\ref{yield} and Eq.~\ref{fluence} are known. Thus, we can consider all possible values of $\beta$ and $A_P$ compatible with the observed muon counting rate excess, $J_{\mu}$, and the integrated GOES-proton fluence $F$. Fig.~\ref{cruze} summarizes the situation. Giving:\\
$Ap=(1.20\pm 0.96) \times 10^{-3}/(cm^2s\;sr\;GeV)$\\ and $\beta=1.89\pm 1.10$.

To make a comparison with satellite GOES data, we obtain the integral proton flux in the GeV energy range as
\begin{equation}
J_P(>E_P)=\int_{Emin}^{\infty} A_P E_P^{-\beta}dE_P.
\end{equation}
Fig.~\ref{tupi_flux} shows the results of the integral proton flux obtained from the New-Tupi muon excess observed in coincidence with the radiation. The red circles represent the GOES-16 data, and the black squares represent the expected proton flux in the GeV energy range obtained from Monte Carlo, muon excess at the New-Tupi detector, and the GOES-proton fluence.
The origin of this transient event was the solar eruption, an M6-class flare (prompt emission), accelerating protons (ions) up to relativistic energies, GeV energy range.

\section{Conclusions}

We have reported evidence of SEPs accelerated up to GeV energies during the eruptive phase of the M6-class solar flare on July 17, 2023. The result comes from a timing analysis of a muon excess from the New-Tupi detector at the SAA central region. Muons at New-Tupi are produced by protons (ions) interaction in the upper atmosphere reaching the Earth with a magnetic rigidity above 3 GV ($\sim$ 3 GeV for protons).
 
In most cases, SEP (mostly protons) detected by the GOES-16 spacecraft shows two steps. An impulsive phase, where the acceleration of protons (ions) is by the prompt emission of flare, followed by a gradual phase of long duration, where the acceleration of protons (ions) is by the associated CME shock waves.

We want to point out that the muon excess produced by SEPs with an effective rigidity above the cutoff ($\sim$ 3 GV) at the New-Tupi muon detector is in temporal coincidence only with the GOES proton flux from the impulsive phase (see Fig.~\ref{goes_tupi}) . Consequently, in the gradual phase, the protons accelerated by CME's shocks do not reach the GeV energy range because a muon excess is absent at ground level.

A marginal muon excess also appears on the Yan ba Jing S-21(pointing 21 degrees south) muon telescope (in Tibet). Also, a marginal particle excess is seen only in the French Kerguel NM (close to the South polar region).
In both cases, the excesses are in (temporal) coincidence with the GOES proton flux (impulsive phase). However, it is hard to verify whether these excesses are genuine due to low confidence or simply fluctuations in the detectors' count rate.

From a Monte Carlo analysis,
we show that the SAA central region is favourable to the observation of transient solar events, especially SEP, because 
the magnetosphere  has a "dip´´ in this region, 
weakening the geomagnetic field strength and allowing the entrance of charged particles at large deeps in a region not far from the geographic Equator, giving a rigidity sub-cutoff around 3.1 GV in a place where the conventional Stormer geomagnetic rigidity cutoff is around 10 GV.

\section{Acknowledgments}

This work is supported by the Rio de Janeiro Research Foundation (FAPERJ) under Grant E-26/010.101128/2018. We thank to NMDB Database (www.nmdb.eu), founded under the European Union FP7 Program (Contract No. 213007)
by provide NMs data and the Space Weather Prediction Center
from NOAA for its open data policy.

\appendix

\section{New-Tupi detector}

The New-Tupi telescope is built with four identical particle detectors, forming two telescopes, as shown in Fig. 8 from \cite{augu16a}. Each detector consisting of an Eljen EJ-208
plastic scintillator slab of 150 cm x 75 cm x 5 cm and a Hamamatsu R877 photomultiplier of 127
millimeters in diameter, packaged in a pyramidal box.
The PMT high voltage divider, amplifier, and high voltage power supplier are in the ORTEC ScintiPackTM Photomultiplier Base 296.

From February 6, 2023, we have implemented a data acquisition system using a VERTILON high-speed pulse counting system (MCPC618-8 Channel). allowing for direct connection with the PMTs without the need for external preamplifiers,
with a 250 MHz count rate per channel.

Now the detector is working only in scaler mode or 
single particle technique \citep{agli96}, where the single hit rates of all four PMTs, are recorded once a second.
 However, so far, only two detectors are working. The coincidences among these detectors of each telescope will be implanted. 
 
Also, the barometric coefficients for cosmic muon fluxes at the Earth's surface can be obtained using the CORSICA code in \cite{kovy13}. For New-Tupi detector conditions and at sea level, the barometric coefficient is about -0.14\% per mb, about eight to nine times less than the typical barometric coefficient in NMs.

\newpage

\bibliography{s2_storm}
\bibliographystyle{aasjournal}

\end{document}